\documentclass[journal]{IEEEtran}
\usepackage{upgreek}
\usepackage{amsmath,amsfonts,amssymb,mathrsfs}
\usepackage{enumerate}
\usepackage{color}
\usepackage{cite}
\usepackage{graphicx}\usepackage{subfigure}
\usepackage{booktabs}
\usepackage{multirow}
\usepackage{cases}
\usepackage{breakurl}
\usepackage{float}
\usepackage{algorithm}
\usepackage[noend]{algpseudocode}
\newlength{\continueindent}
\usepackage{etoolbox}
\makeatletter
\newcommand*{\ALG@customparshape}{\parshape 2 \leftmargin \linewidth \dimexpr\ALG@tlm+\continueindent\relax \dimexpr\linewidth+\leftmargin-\ALG@tlm-\continueindent\relax}
\apptocmd{\ALG@beginblock}{\ALG@customparshape}{}{\errmessage{failed to patch}}
\makeatother
\makeatletter
\newcommand{\multiline}[1]{%
  \begin{tabularx}{\dimexpr\linewidth-\ALG@thistlm}[t]{@{}X@{}}
    #1
  \end{tabularx}
}
\makeatother

\usepackage{extarrows}
\makeatletter
\newcommand{\removelatexerror}{\let\@latex@error\@gobble}
\makeatother

\newcommand{\etal}{\textit{et al.} }
\newcommand{\equref}[1]{(\ref{#1})}
\newcommand{\figref}[1]{Fig.~\ref{#1}}
\newcommand{\tabref}[1]{Table~\ref{#1}}
\newcommand{\secref}[1]{Section~\ref{#1}}

\newcommand{\textdssim}{{D}_\text{SSIM}}
\newcommand{\textdmse}{{D}_\text{MSE}}

\newcommand{\dssim}{D_\text{SSIM}}
\newcommand{\dmse}{D_\text{MSE}}

\newcommand{\lssim}{{\lambda_\text{SSIM}}}
\newcommand{\lmse}{{\lambda_\text{MSE}}}

\begin{document}
\title{SSIM-Based CTU-Level Joint Optimal Bit Allocation and Rate Distortion Optimization}

\author{Yang~Li
        and~Xuanqin~Mou,~\IEEEmembership{Senior~Member,~IEEE}
\thanks{Yang~Li and Xuanqin~Mou are with the Institute of Image Processing and Pattern Recognition, Xi'an Jiaotong University, Xi'an 710049, China (e-mail: liyang2012@stu.xjtu.edu.cn; xqmou@mail.xjtu.edu.cn).}
}
\maketitle
\begin{abstract}
Structural similarity (SSIM)-based distortion $\dssim$ is more consistent with human perception than the traditional mean squared error $\dmse$.
To achieve better video quality,
many studies on optimal bit allocation (OBA) and rate-distortion optimization (RDO) used $\dssim$ as the distortion metric.
However, many of them failed to optimize OBA and RDO jointly based on SSIM, thus causing a non-optimal R-$\dssim$ performance.
This problem is due to the lack of an accurate R-$\dssim$ model that can be used uniformly in both OBA and RDO.
To solve this problem,
we propose a $\dssim$-$\dmse$ model first.
Based on this model, the complex R-$\dssim$ cost in RDO can be calculated as simpler R-$\dmse$ cost with a new SSIM-related Lagrange multiplier.
This not only reduces the computation burden of SSIM-based RDO, but also enables the R-$\dssim$ model to be uniformly used in OBA and RDO.
Moreover, with the new SSIM-related Lagrange multiplier in hand, the joint relationship of R-$\dssim$-$\lssim$ (the negative derivative of R-$\dssim$) can be built, based on which the R-$\dssim$ model parameters can be calculated accurately.
With accurate and unified R-$\dssim$ model, SSIM-based OBA and SSIM-based RDO are unified together in our scheme, called SOSR.
Compared with the HEVC reference encoder HM16.20,
SOSR saves 4\%, 10\%, and 14\% bitrate
under the same SSIM in all-intra, hierarchical and non-hierarchical low-delay-B configurations, which is superior to other state-of-the-art schemes.
An improved version of this manuscript has been accepted by IEEE Transactions on Broadcasting DOI 10.1109/TBC.2021.3068871.
The project page is located at http://gr.xjtu.edu.cn/web/xqmou/sosr.
\end{abstract}

\IEEEpeerreviewmaketitle

\section{Introduction}
High Efficiency Video Coding (HEVC) standard has achieved significant compression performance improvement compared with the previous H.264 standard \cite{Sullivan2012}.
However, due to the widely available applications such as video on demanding, video streaming and video chatting, the burden of video transmission and storage is still growing.
Faced with this situation,
how to control the encoding to achieve the minimum possible distortion with the limited bits becomes a fundamental challenge.

For HEVC, the encoding is controlled by many encoding parameters (e.g., quantization parameter (QP) and Lagrange multiplier $\lambda$), as well as a large number of encoding modes (e.g., block partition mode and prediction mode).
Thus, the encoding controlling is actually a problem of encoding parameters and modes selection \cite{wiegand2003rate}.
In practice,
aiming at achieving the minimum distortion with the limited bits,
the encoder can select the best combination of parameters and modes in three steps.
First,
the encoder determines how many bits should be allocated to every encoding units (e.g., all the coding tree units (CTUs) in a frame) to produce minimal distortion, known as optimal bit allocation (OBA).
Secondly,
the appropriate encoding parameters such as QP and $\lambda$ are determined to achieve the bits allocated to a unit in the first step, which is called bitrate control.
Thirdly, applying the determined parameters,
the encoder traverses all possible modes to encode the unit, and the mode with the least R-D cost (i.e., D+$\lambda$R) is selected as the best mode, known as rate distortion optimization (RDO).

In the three steps,
OBA can be solved using the Lagrangian optimization by modeling the R-D property of each unit \cite{sullivan1998rate,ortega1998rate}.
The R-D property is the relationship between the bits consumed and the distortion produced.
By introducing the Lagrangian multiplier $\lambda$,
the objective of OBA is equivalent to minimizing the total R-D cost of all the units.
Specifically, $\lambda$ is the negative derivative of R-D.
Thus, an R-$\lambda$ relationship can be obtained from the R-D model, which can be used to determine a proper $\lambda$ associated with the bits allocated by OBA. Moreover, $\lambda$ is an encoding parameter, which balances bits and distortion in the R-D cost of RDO.
Therefore, OBA, bitrate control, and RDO can be unified together into finding a proper $\lambda$ to satisfy the bits constraint.
Such a $\lambda$-based scheme proposed in \cite{li2012rate,li2013adaptive} has been adopted by the HEVC reference encoder HM \cite{hmweb}.

Besides,
the Lagrangian optimization relies on an accurate model of the R-D property.
However, before a unit is encoded, its R-D property is unknown.
Typically, the R-D property can be modeled by the exponential \cite{hang1997source} or hyperbolic \cite{kamaci2005frame} functions with two unknown parameters.
Many studies adopt a statistical regression method to estimate the model parameters based on a set of R-D data collected from previously encoded units \cite{Chiang1997,Choi2013,li2012rate}.
Using this method, the model parameters of a series of collocated units will always be similar even in scenes with fast motion \cite{Gao2017}.
Thereby, its estimation accuracy is not satisfactory.
To overcome this limitation, there are three common methods to be used.
First, learning-based approach can be used to predict the model parameters of a unit before encoding, such as in \cite{Gao2017,li2017convolutional,JVETM0215,santamaria2018estimation}.
Secondly, two-pass encoding such as in \cite{Fiengo2017} and \cite{Zupancic2017} is also effective to estimate the model parameters by using the statistics in the first-pass encoding as a priori.
Thirdly,
since $\lambda$ can be obtained by deriving R-D, the R-D-$\lambda$ constitute a joint relationship containing only two unknown parameters, which can be uniquely solved with the encoding results of a unit
without relying on a series of units as in the regression method \cite{Li2017}.
The three methods have been verified to be more accurate in parameter estimation than the regression method.
Accordingly, the R-D property can be better characterized and the corresponding studies have achieved satisfied R-D performance.

In the studies mentioned above, mean squared error (MSE, the variable is denoted by $\dmse$) is usually applied as the distortion metric, which measures the pixel-wise difference between the encoded and the original videos. However, MSE has been validated to be poorly correlated with the human subjective perception of distortion [30].
Thus, minimizing $\dmse$ of the encoded video cannot achieve an optimal perceptual quality.
To overcome this problem, the influential perceptual quality metrics such as the well-known Structural SIMilarity index (SSIM) [31] have been adopted in many recent studies on OBA and RDO. SSIM evaluates the similarity of luminance, contrast, and structures between two images, to which human perception is highly sensitive.
Thus, SSIM has better consistency with human perception than MSE.
SSIM ranges from 0 to 1, with larger values indicating better quality.
Thus, SSIM-based distortion denoted by $\dssim$ can be calculated as 1-SSIM.
By using $\dssim$ as the distortion metric for encoding optimization,
better perceptual quality has been achieved in the studies such as
\cite{Ou2011,Gao2016,Zhou2019,Wang2015_twopass,Wang2012ssim,wang2013perceptual,Yeo2013,Dai2014ssim,qi2013efficient,yeo2013ssim}.

However, the adoption of SSIM has also brought new problems.
Although there have been a lot of SSIM-based OBA studies (e.g., \cite{Ou2011,Gao2016,Zhou2019}),
their bitrate control and RDO were still based on MSE.
Accordingly, the mode that has the least R-$\dmse$ cost ($\dmse\!+\!\lmse R$) will be selected.
The corresponding R-$\dssim$ cost ($\dssim\!+\!\lssim R$) is not guaranteed to be minimal.
On the other hand, there are also many SSIM-based RDO schemes \cite{Yeo2013,Wang2012ssim,wang2013perceptual,Dai2014ssim}.
However, the SSIM-based OBA was not studied in these works.
That is, in these SSIM-based studies, OBA and RDO are optimized based on the R-$\dssim$ model and the R-$\dmse$ model, respectively, or vice versa.
Hence, the resulting encoding is not optimal in the R-$\dssim$ performance with considering the performance of both OBA and RDO.

To achieve better R-$\dssim$ performance, an accurate R-$\dssim$ model should be uniformly used in both OBA and RDO, but there are two problems that prevent this approach.
First, calculating R-$\dssim$ cost for RDO is too time-consuming. Specifically, the time cost of $\dssim$ is 10 times of that of $\dmse$ \cite{xue2013perceptual}.
Since HEVC has a large number of possible modes to encode a CTU \cite{Sullivan2012},
the high complexity of $\dssim$ will bring a huge increase in mode decision time of RDO.
This is why these SSIM-based OBA studies \cite{Ou2011,Gao2016,Zhou2019} used MSE-based RDO, although MSE-based RDO is not bound to achieve optimal R-$\dssim$ performance.
To solve this problem, some studies on SSIM-based RDO built a model between $\dssim$ and $\dmse$ (e.g., \cite{Yeo2013,Dai2014ssim,qi2013efficient,yeo2013ssim}).
Based on this model, the R-$\dssim$ cost is equivalent to a modified R-$\dmse$ cost. 
Thus, the calculation of $\dssim$ is avoided during RDO that saves the encoding time.
However, 
accuracy of their $\dssim$-$\dmse$ model is less than satisfactory.
This is why the corresponding R-$\dssim$ model was not used to solve SSIM-based OBA in their work, and thus accurate R-$\dssim$ modeling is the second problem.

In fact,
even for many SSIM-based OBA studies,
accurate R-$\dssim$ modeling remains a challenging task.
We can see that traditional regression method has still been commonly used to estimate the R-$\dssim$ model parameters such as in \cite{Ou2011,Gao2016,Zhou2019}, which leaves room for further improvement in estimation accuracy.
There are a few of learning-based methods presenting promising performance in R-$\dssim$ modeling for images \cite{Xu2017}. However, their effectiveness in OBA for video encoding needs further verification.
The second method, i.e., the two-pass encoding, also shows improved accuracy of the R-$\dssim$ model parameter estimation \cite{Wang2015_twopass}.
However, it brings twice the encoding time.
Alternatively, the R-D-$\lambda$ joint relationship-based method is more accurate and has similar complexity compared with the traditional regression method, and hence it can be used to achieve more accurate R-$\dssim$ model.
However, if RDO is not based on SSIM, the $\lssim$ associated with the resulting R and $\dssim$ is unknown. Hence, the joint relationship cannot be solved.

{ In this study, we try to solve the two problems.
First,
an accurate $\dssim$-$\dmse$ model is proposed, based on which
SSIM-based RDO can be performed with the simpler R-$\dmse$ cost, thus avoiding the increase in encoding time. 
At the same time, based on this model, we also established the model between $\lssim$ and $\lmse$, so that $\lssim$ calculated by SSIM-based OBA can be used in SSIM-based RDO, which makes the R-$\dssim$ model in OBA and RDO unified.
Moreover,
by using $\lssim$ in the RDO process,
the association between $\lssim$ and the resulting R and $\dssim$ can be built.
Accordingly, the joint R-$\dssim$-$\lssim$ relationship can be exploited to calculate the R-$\dssim$ model parameters accurately.
With accurate and unified R-DSSIM model,
SSIM-based OBA and SSIM-based RDO are unified together in
our scheme.
Experimental results verified that the proposed scheme achieves better R-$\dssim$ performance than the state-of-the-art OBA schemes in All-Intra (AI), hierarchical, and non-hierarchical Low-delay-B (h\_LB/nh\_LB) configurations.
The main contributions are summarized as follows:
\begin{itemize}
\item A new $\dssim$-$\dmse$ model is proposed.
    Compared with the widely used $\dssim$-$\dmse$ model as in \cite{Yeo2013,Dai2014ssim,qi2013efficient,yeo2013ssim}, the proposed model is more accurate.
    Moreover, based on this model, we unify the R-$\dssim$ model in OBA and RDO.
\item The joint R-$\dssim$-$\lssim$ relationship is proposed to calculate the R-$\dssim$ model parameters.
    It is worth noting that without the unification of R-$\dssim$ model in OBA and RDO as proposed in this study, this joint relationship cannot be solved.
\item With accurate R-$\dssim$ model uniformly used in OBA and RDO,  {S}SIM-based {O}BA and SSIM-based {R}DO scheme are unified together in the proposed scheme  called SOSR, achieving an outstanding R-$\dssim$ performance.
\end{itemize}

The rest of the paper is organized as follows.
Section~II introduces the background.
In Section~II, drawbacks of the conventional scheme are analyzed.
Section~IV describes the proposed scheme in detail.
Section~V presents the experiments to validate the effectiveness of the proposed scheme.
Finally, Section~VI concludes this paper.

\section{Backround}

\subsection{SSIM}
SSIM measures the degradation of structural information in the distorted image $\mathbf{y}$ compared to the pristine image $\mathbf{x}$.
(In video coding, $\mathbf{x}$ and $\mathbf{y}$ are the original and reconstructed frames, respectively.)
Specifically, a SSIM map is calculated first by comparing the luminance similarity, contrast similarity, and structural similarity between $\mathbf{x}$ and $\mathbf{y}$.
Then, mean of the SSIM map is calculated as the overall SSIM index of $\mathbf{y}$.
The standard SSIM map \cite{Wang2004} is defined as follows:
\begin{equation}\label{equ:ssimmap}
  \text{SSIMmap} = \frac{2\mu_\mathbf{x} \mu_\mathbf{y} + C_1}{\mu_\mathbf{x}^2+\mu_\mathbf{y}^2+C_1} \cdot \frac{2\sigma_{\mathbf{xy}}^2+C_2}{\sigma_{\mathbf{x}}^2 + \sigma_{\mathbf{y}}^2+C_2},
\end{equation}
where $\mu$ and $\sigma$ denote the local mean and local variance, respectively. $\sigma_\mathbf{xy}$ is the local covariance of $\mathbf{x}$ and $\mathbf{y}$. These local operations are calculated based on a $11\times11$ Gaussian weighted block around each pixel \cite{Wang2004}. Besides, $C_1$ and $C_2$ are constants to prevent dividing by zero.

SSIM is a quality index ranging from 0 to 1, with larger values indicating better quality.
Thus, the SSIM-based distortion of a unit (a frame or a CTU) can be calculated as:
\begin{equation}\label{equ:dssim}
  \dssim=1-\dfrac{1}{\text{M}} \sum_{p\in{\text{unit}}} \text{SSIMmap}(p),
\end{equation}
where $p$ denotes the pixel in SSIM map belonging to the unit and $\text{M}$ is the total number of $p$.
Based on this calculation,
the relationship of $\dssim$ between a frame and CTUs in the frame can be formulated as follows:
\begin{equation}\label{equ:dframe}
  {{\dssim}_\text{frame}}_j = \dfrac{1}{\text{M}_\text{frame}}\sum_{i=0}^{N-1} \left({\dssim}_{i,j}\cdot\text{M}_{i,j}\right),
\end{equation}
where ${{\dssim}_\text{frame}}_j$ indicates the distortion of the $j$-th frame and ${\dssim}_{i,j}$ indicates the distortion of the $i$-th CTU in the frame (denoted by $\text{CTU}_{i,j}$).
$\text{M}_\text{frame}$ and $\text{M}_{i,j}$ are the number of pixels in the frame and in the CTU, respectively, and $N$ is the total number of CTUs in the frame.

\subsection{SSIM-based OBA schemes}
The SSIM-based OBA can be formulated as follows:
\begin{equation}\label{equ:oba}
  \mathop{\arg\min}_{\{R_{i,j}\}_{i=0}^{N-1}} {{\dssim}_\text{frame}}_j,\
  \text{s.t.,} \sum_{i=0}^{N-1} R_{i,j}\leq{R_c},
\end{equation}
where $R_{i,j}$ is the allocated bits for $\text{CTU}_{i,j}$, which needs to be solved.
Typically, distortion is regarded as the function of the allocated bits in OBA.
For example, Ou \etal \cite{Ou2011} used the exponential function to model the R-$\dssim$ for H.264.
Then, the Lagrangian multiplier method was used to calculate the optimal allocated bits.
In \cite{Gao2016}, Gao \etal expressed the R-$\dssim$ model as a hyperbolic function.
The SSIM-based OBA for intra encoding of HEVC was then by the bargaining game-based optimization.
In \cite{Zhou2019}, $\dssim$ was calculated according to the divisive normalization theory. The corresponding R-$\dssim$ was modeled as a logarithmic function.
For the model parameters estimation, the traditional regression method was still used in \cite{Ou2011,Zhou2019}.
To improve the R-$\dssim$ modeling accuracy,
a convolutional neural network was proposed in \cite{Xu2017} that predicts SSIM and bits of an image encoded with different QPs.
Better R-$\dssim$ modeling accuracy has been achieved than the traditional regression-based strategy.
In \cite{Wang2015_twopass}, Wang \etal proposed a SSIM-based two-pass encoding.
The statistics based on the encoding results in the first-pass were exploited to improve the R-$\dssim$ modeling accuracy.
It is worth noting that the RDO in these studies is still performed based on MSE, which has not yet been optimal to maximize the R-$\dssim$ performance.

\subsection{Bitrate Control}
Bitrate control is performed to determine the encoding parameters used in RDO aiming to achieve the allocated bits for each unit.
QP is used to be the key encoding parameter.
Early studies mainly built the R-QP model for bitrate control, such as the widely used quadratic R-QP model \cite{Chiang1997,Choi2013}.
For HEVC, the model between R and the Lagrangian multiplier $\lambda$ has been verified to be more effective in bitrate control than the R-QP model
 \cite{li2012rate,karczewicz2013intra,li2013adaptive}.
Particularly,
$\lambda$ is an encoding parameter of RDO that balances the tradeoff between R and D. QP can also be determined by a QP-$\lambda$ model before RDO \cite{li2012qp}.
Moreover, the R-$\lambda$ model, which can be built as the derivative of R-D, reflects the R-D property.
Therefore, OBA, bitrate control, and RDO can be unified together into searching the optimal $\lambda$
that is associated with the constrained bits.
However, these studies are all based on MSE.

\subsection{SSIM-based RDO schemes}
RDO is a process to select the encoding mode that has the minimal R-D cost.
For SSIM-based RDO, it is natural to calculate the R-$\dssim$ cost to replace the traditional R-$\dmse$ cost.
Many studies such as \cite{mai2005new,yang2009ssim,Huang2010} used this approach.
However, this approach needs to calculate $\dssim$ for each candidate mode.
Although the calculation of SSIM can be accelerated with the divisive normalization theory as in \cite{Wang2012ssim,wang2013perceptual},
the encoding time will still be increased a lot compared to that based on MSE.
To reduce the computation burden,
Yeo \etal \cite{Yeo2013} approximated $\dssim$ to $\dmse$ that is normalized by the local variance of the block.
This method was widely adopted in many studies such as \cite{Dai2014ssim,qi2013efficient,yeo2013ssim}.
In this way, the R-$\dssim$ cost is equal to the simpler R-$\dmse$ cost with a scaled $\lssim$.
However, based on our evaluation, accuracy of their model is less than satisfactory.

\section{Problem Analysis}\label{sec:problem}
As has been discussed in the introduction section, many studies such as \cite{Ou2011,Gao2016,Zhou2019}
proposed to combine the SSIM-based OBA and MSE-based RDO (named as SOMR in this study).
To analyze drawbacks of this kind of scheme,
this section proposes a typical implementation of SOMR, which is summarized in Algorithm~1.

Specifically, according to the Lagrange multiplier method, the SSIM-based OBA problem as in \equref{equ:oba} is equivalent to the unconstrained problem as follows:
\begin{equation}\label{equ:rdo}
  \begin{aligned}
  &\mathop{\arg\min}_{\{R_{i,j}\}_{i=0}^{N-1}} J =  \sum_{i=0}^{N-1} \left({\dssim}_{i,j}\cdot\text{M}_{i,j}\right) + \lssim\cdot \sum_{i=0}^{N-1} R_{i,j}\\
  =&\mathop{\arg\min}_{\{bpp_{i,j}\}_{i=0}^{N-1}}J=  \sum_{i=0}^{N-1} \text{M}_{i,j} \cdot \left( {\dssim}_{i,j} + \lssim\cdot {bpp}_{i,j} \right)\\
  \end{aligned}
\end{equation}
where $bpp_{i,j}$ denotes bits per pixel, i.e., $bpp_{i,j}=R_{i,j}/M_{i,j}$. The optimal solution of \equref{equ:rdo} can only be reached when
the bit constraint is satisfied
\begin{equation}\label{equ:bitsconstraints}
\sum_{i=0}^{N-1} R_{i,j} = R_c
\end{equation}
and
\begin{equation}\label{equ:J}
  \dfrac{\partial{J}}{\partial{bpp_{i,j}}}=\dfrac{\partial{{\dssim}_{i,j}}}{\partial{bpp_{i,j}}}+\lambda_\text{SSIM}\!=\!0.
\end{equation}
\equref{equ:J} explains that $\lssim$ is the negative derivative of R-$\dssim$.

\begin{table}[]
\centering
\caption{Hyperbolic R-$\dssim$ modeling performance in terms of correlation coefficient for videos encoded in AI and LB configurations.}
\label{tab:rdssimcc}
\begin{tabular}{ccccccc}
\toprule
class&A      & B      & C      & D      & E      & avg.           \\
\midrule
AI & 0.9537 & 0.9287 & 0.9548 & 0.9640 & 0.8838 & 0.9377 \\
LB & 0.9908 & 0.9761 & 0.9898 & 0.9866 & 0.9277 & 0.9750\\
\bottomrule
\end{tabular}
\end{table}

To solve \equref{equ:J}, we adopt the widely used hyperbolic function as the R-$\dssim$ model, which is expressed as
\begin{equation}
{\dssim}_{i,j} = \alpha_{i,j} \cdot {bpp_{i,j}}^{\beta_{i,j}},
\label{equ:rdmodel}
\end{equation}
where $\alpha_{i,j}$ and $\beta_{i,j}$ are the unknown model parameters.
To verify the effectiveness of this model, we encoded five classes of video sequences with four different QPs \{22,27,32,37\} in AI and LB configurations, separately.
The hyperbolic R-$\dssim$ model was fitted based on the resulted R and $\dssim$ at four QPs for each CTU.
Correlation coefficient between the actual $\dssim$ and the predicted ones by the fitted model is then calculated.
\tabref{tab:rdssimcc} lists the average correlation coefficient results, which verify the high accuracy of the hyperbolic R-$\dssim$ model.

Based on \equref{equ:J} and \equref{equ:rdmodel}, the R-$\lssim$ model can be obtained
\begin{equation}\label{equ:rlambda}
{\lssim}_{i,j}  = -\alpha_{i,j} \beta_{i,j} \cdot {bpp}_{i,j}^{\beta_{i,j} -1}.
\end{equation}
By substituting \equref{equ:rlambda} into \equref{equ:bitsconstraints}, we can search the optimal $\lssim$ with the Bisection method to meet the bits constraint, that is,
\begin{equation}\label{equ:optimallambda}
  \sum_{i=0}^{N-1} \text{M}_{i,j}\cdot \left(\dfrac{{\lambda^*_\text{SSIM}}_{i,j}}{-\alpha_{i,j} \beta_{i,j}}\right)^{\frac{1}{\beta_{i,j}-1}} = \sum_{i=0}^{N-1} R^*_{i,j} =   R_c,
\end{equation}
where $R^*_{i,j}$ is the optimal bits allocated to $\text{CTU}_{i,j}$ and
${\lambda^*_\text{SSIM}}_{i,j}$ is the corresponding optimal $\lssim$.
With ${\lambda^*_\text{SSIM}}_{i,j}$ in hand, by traversing all possible modes to encode $\text{CTU}_{i,j}$,
the mode that has the least R-$\dssim$ cost can be set as the optimal mode.

However,
as has been discussed,
many SSIM-based OBA studies, e.g., \cite{Ou2011,Gao2016,Zhou2019}, adopted the default bitrate control and RDO scheme of HM
that is based on MSE.
Specifically, taking the inter encoding as an example, HM uses an R-$\lmse$ model \cite{li2013adaptive} to calculate the $\lambda_\text{MSE}$ associated with $R^*_{i,j}$ for each CTU, i.e.,
\begin{equation}\label{equ:add_rl}
\lmse_{i,j}=c_{i,j} \left(\frac{R^*_{i,j}}{M_{i,j}}\right)^{k_{i,j}}
\end{equation}
$\lmse_{i,j}$ is then used in the calculation of R-$\dmse$ cost during the MSE-based RDO.

Besides, calculating $\lmse_{i,j}$ needs to estimate the unknown R-$\lmse$ model parameters $c_{i,j}$ and $k_{i,j}$.
In \cite{li2013adaptive}, after encoding each CTU, the R-$\lmse$ model parameters are estimated in a regression method
for the subsequent frames as follows:
\begin{equation}\label{equ:updatingrlambda}
  \left\{
  \begin{aligned}
    &\Delta {\lmse}_{i,j} =\ln {\lmse}_{i,j}\!-\!\ln{(c_{i,j} {{bpp}_{i,j})}^{k_{i,j}}}\\
    &c_{i,j+1} \!=\!c_{i,j}\!+\!\delta_c\! \cdot\! \Delta {\lmse}_{i,j} \cdot c_{i,j} \\
    &k_{i,j+1}  \!=\!k_{i,j}\!+\!\delta_k\!\cdot\! \Delta {\lmse}_{i,j} \cdot \ln{{bpp}_{i,j}},
  \end{aligned}
  \right.
\end{equation}
where $\delta_c$ and $\delta_k$ control the updating speed.
Inspired by this method,
the unknown R-$\dssim$ model parameters are estimated in a similar way in the proposed SOMR scheme:
\begin{equation}\label{equ:updatingrdssim}
  \left\{
  \begin{aligned}
    &\Delta {\dssim}_{i,j} =\ln {\dssim}_{i,j}\!-\!\ln{(\alpha_{i,j} {{bpp}_{i,j})}^{\beta_{i,j}}}\\
    &\alpha_{i,j+1} \!=\!\alpha_{i,j}\!+\!\delta_\alpha\! \cdot\! \Delta {\dssim}_{i,j} \cdot \alpha_{i,j} \\
    &\beta_{i,j+1}  \!=\!\beta_{i,j}\!+\!\delta_\beta\!\cdot\! \Delta {\dssim}_{i,j} \cdot \ln{{bpp}_{i,j}}.
  \end{aligned}
  \right.
\end{equation}
The regression-based parameter estimation method is also used in many SSIM-based studies such as \cite{Ou2011,Zhou2019}.

\renewcommand{\algorithmicrequire}{\textbf{Input:}}
\renewcommand{\algorithmicprocedure}{{Step}}
\begin{algorithm}[t]
\caption{SOMR: SSIM-based {O}BA and MSE-based {R}DO}
    \begin{algorithmic}
       \Require{$\alpha_{i,j}$, $\beta_{i,j}$, $\theta_{i,j}$, $i\in\{0,1,\cdots N\}$}
       \Procedure{1: \textit{SSIM-Based Optimal Bit Allocation}}{}
       \State Search the optimal ${\lambda^*_\text{SSIM}}_{i,j}$ that satisfies \equref{equ:optimallambda} with the Bisection method;
       \State Record the corresponding $R^*_{i,j}$;
       \EndProcedure

       \Procedure{2: \textit{CTU Encoding With MSE-Based RDO}}{}
       \For{$i=0:N-1$}
       \State a.  MSE-Based Bitrate Control:
       \State \hskip2.5em Calculate $\lmse_{i,j}$ by \equref{equ:add_rl};
       \State b. Traverse all possible encoding modes;
       \State c. Search mode with minimum R-$\dmse$ cost;
       \State d. Encode  $\text{CTU}_{i,j}$ with the selected mode;
       \EndFor
       \EndProcedure

       \Procedure{3: \textit{Parameter Estimation For Subsequent Frame}}{}
       \For{$i=0:N-1$}
       \State a. Record $bpp_{i,j}$, ${\dssim}_{i,j}$ and, ${\lambda_\text{MSE}}_{i,j}$;
       \State b. Update R-$\lmse$ model parameters by \equref{equ:updatingrlambda};
       \State c. Update R-$\dssim$ model parameters by \equref{equ:updatingrdssim};
       \EndFor
       \EndProcedure
    \end{algorithmic}
\end{algorithm}

\begin{table}[]
\centering
\caption{R-$\dssim$ performance of SOMR.}
\label{tab:SOMR}
\resizebox{\linewidth}{!}
{
\begin{tabular}{ccccccc}
\toprule
      & \multicolumn{2}{c}{AI} & \multicolumn{2}{c}{h\_LB} & \multicolumn{2}{c}{nh\_LB} \\
\cmidrule(lr){2-3}
\cmidrule(lr){4-5}
\cmidrule(lr){6-7}
class & BDBR       & BD-SSIM   & BDBR        & BD-SSIM     & BDBR         & BD-SSIM     \\
\midrule
A     & -1.90\%    & 0.0001    & -7.60\%     & 0.0005      & -11.10\%     & 0.0007      \\
B     & -0.10\%    & 0.0000    & -5.80\%     & 0.0014      & -10.10\%     & 0.0018      \\
C     & 1.10\%     & -0.0005   & -2.00\%     & 0.0006      & -3.00\%      & 0.0012      \\
D     & -0.10\%    & 0.0001    & -3.40\%     & 0.0020      & -4.30\%      & 0.0030      \\
E     & -2.30\%    & 0.0002    & 0.70\%      & 0.0000      & -0.50\%      & 0.0000      \\
avg.  & -0.70\%    & 0.0000    & -3.50\%     & 0.0010      & -5.70\%      & 0.0015      \\
\bottomrule
\end{tabular}
}
\end{table}

To evaluate the R-$\dssim$ performance of SOMR,
five classes of video sequences were encoded at four different QPs \{22,27,32,37\} respectively by SOMR and the anchor schemes, which are the default MSE-based OBA and RDO schemes of HM16.20, i.e., JCTVC-M0257 \cite{karczewicz2013intra} for AI and JCTVC-M0036 \cite{li2013adaptive} for LB.
Then,
Bj{\o}ntegaard delta bitrate (BDBR) and the BD-SSIM \cite{bdrate} are evaluated.
BDBR calculates the relative increase of bitrate of SOMR at the same SSIM compared with the anchor, while BD-SSIM calculates the absolute SSIM gain at the same bitrate.
The negative BDBR and positive BD-SSIM indicate the improved R-$\dssim$ performance.
\tabref{tab:SOMR} reports
that SOMR respectively achieves 0.7\%, 3.5\% and 5.7\% BDBR saving in three configurations.
Thus, SOMR does bring some R-$\dssim$ performance improvement. However, the improvement is not significant.
This may be due to two drawbacks of SOMR.
First,
R-$\dssim$ model is not uniformly used in both OBA and RDO of SOMR.
Specifically, SOMR adopts RDO that is based on MSE, with which the encoding mode with the least R-$\dmse$ cost will be selected. Accordingly, the resulted R-$\dssim$ performance is not guaranteed to be optimal.
Secondly, regression-based parameter estimation method has low estimation accuracy.
With inaccurate R-$\dssim$ model parameters, the calculated allocation bits will be non-optimal.
Therefore, it is reasonable to expect that solving these two drawbacks can improve the R-$\dssim$ performance.
However, to the best of our knowledge, no studies have been proposed to solve these two drawbacks.

\begin{figure*}[htp]
    \centering
    \includegraphics[width=\linewidth]{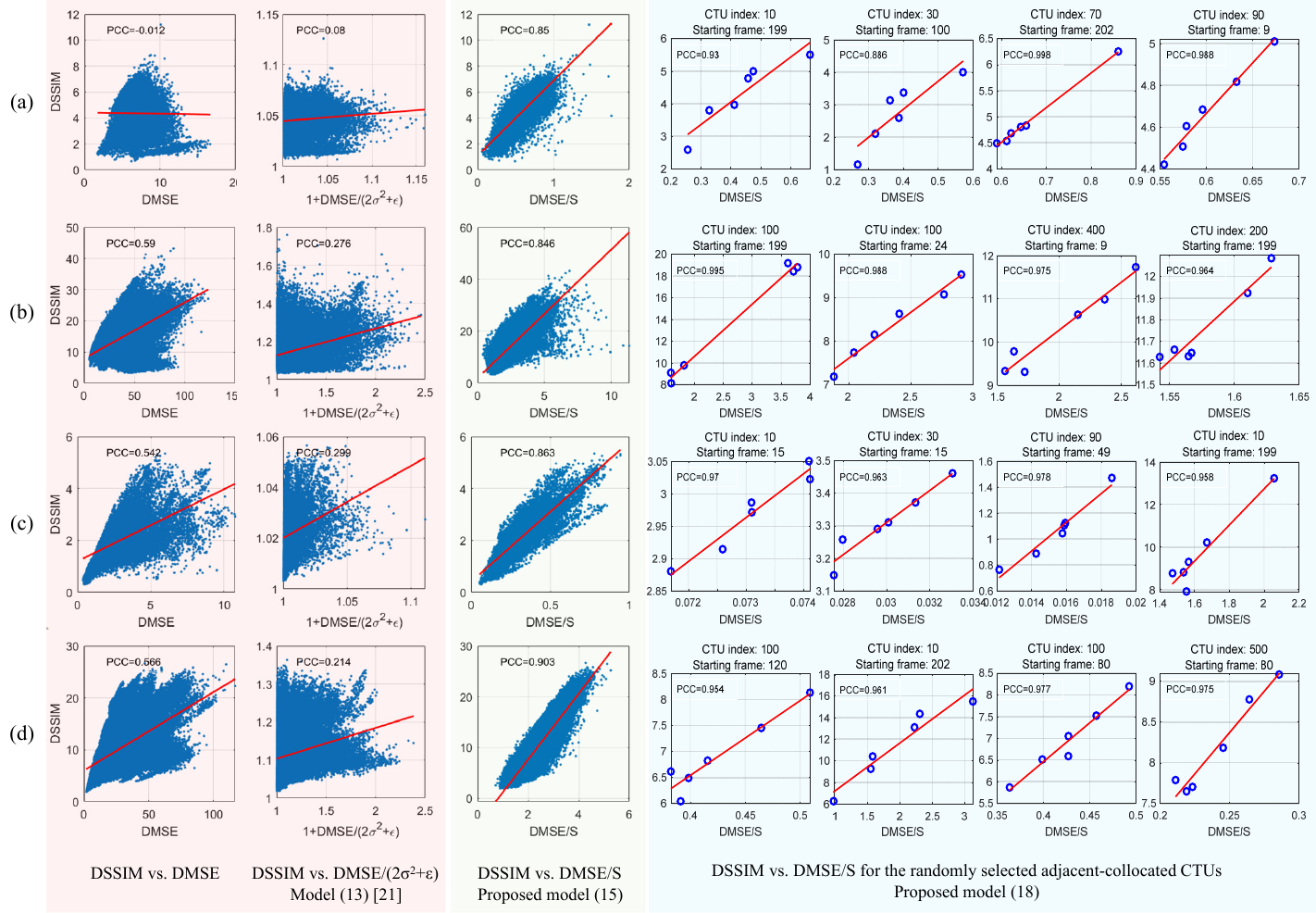}
    \caption{Illustration of the $\dssim$-$\dmse$ relationship. The first column shows the actual values of $\dssim$ vs. $\dmse$. The second and the third column show the $\dssim$-$\dmse$ mapping based on \equref{equ:yeo} \cite{Yeo2013} and the proposed model \equref{equ:globaldd}. The other columns show the $\dssim$-$\dmse$ mapping for six adjacently allocated CTUs that are randomly selected. Our $\dssim$ is multiplied by 100 for better illustration.
    Figures from row (a) to row (d) are generated respectively based on the encoding results of four randomly selected videos: \{Basketballdrill, 480p, LB, QP=22\}, \{Cactus, 1080p, LB, QP=32\}, \{Basketballpass, 240p, AI, QP=22\}, \{Traffic, 1600p, AI, QP=37\}.
    }
    \label{fig:ddrelation}
\end{figure*}

\section{Proposed Joint Optimization Scheme}
This section proposed the joint optimization scheme of SSIM-based OBA and SSIM-based RDO (called SOSR).
Like SOMR, SOSR solves SSIM-based OBA according to \equref{equ:rdo} to \equref{equ:optimallambda} in \secref{sec:problem}.
But the difference is that SOSR solves two problems in SOMR, that is, realizes the unification of R-$\dssim$ model used in OBA and RDO, and improves the modeling accuracy of R-$\dssim$.

\subsection{Solving Drawbacks of SOMR}

For solving the first problem,
R-$\dssim$ cost should be minimized in RDO based on
the optimal ${\lambda^*_\text{SSIM}}_{i,j}$ calculated by SSIM-based OBA \equref{equ:optimallambda}.
At the same time, the high computing complexity of $\dssim$ that will bring intolerant time burden for RDO should be reduced.
For solving the second problem, the joint R-D-$\lambda$ relationship-based parameter estimation method is an effective solution, considering its low complexity and high accuracy.
However,
if $\lssim$ is not used as the encoding parameter in RDO, the association between $\lssim$ and the encoding resulted R and $\dssim$ does not exist. Then, the joint R-D-$\lambda$ relationship cannot be solved.
In this section, we propose a $\dssim$-$\dmse$ model to map the R-$\dssim$ cost to the R-$\dmse$ cost that has a lower complexity, while RDO is still controlled by $\lssim$. In this way, SSIM-based OBA and SSIM-based RDO can be unified.

Specifically, based on a
$\dssim$-$\dmse$ model, the R-$\dssim$ cost can be equivalent to a modified R-$\dmse$ cost, i.e.,
\begin{equation}\label{equ:costmap}
  \begin{aligned}
   & \mathop{\arg\min}_{m}  {\dssim}_{i,j} (m) \cdot\text{M}_{i,j}+{\lambda^*_\text{SSIM}}_{i,j} \cdot R_{i,j}(m)\\
  =& \mathop{\arg\min}_{m}  f({\dmse}_{i,j})(m) \cdot\text{M}_{i,j}+{\lambda^*_\text{SSIM}}_{i,j}\cdot R_{i,j}(m)\\
  \end{aligned}
\end{equation}
where $f$ denotes the $\dssim$-$\dmse$ model.
Based on \equref{equ:costmap}, the time-consuming calculation of $\dssim$ in RDO process can be avoided, thus saving the encoding time.
At the same time,
since ${\lambda^*_\text{SSIM}}_{i,j}$ is still used as the encoding parameter, there is a causal relationship between ${\lambda^*_\text{SSIM}}_{i,j}$ and the $R_{i,j}$ and ${\dssim}_{i,j}$ generated by encoding. Therefore, the R-$\dssim$-$\lssim$ joint relationship is established and can be solved.

However, modeling $\dssim$-$\dmse$ is not an easy problem.
In the first column of \figref{fig:ddrelation}, we illustrate the actual values of $\dssim$-$\dmse$ for four example videos that were encoded by HM16.20 in the default AI or LB configurations. We can see that there is no evident one-to-one mapping between $\dssim$ and $\dmse$.
This is because that $\dssim$ captures the structural degradation, whereas $\dmse$ calculates the pixel-wise error.
Thus, the $\dssim$-$\dmse$ relationship depends on the image content, which varies from region to region.
Basically, the more complex regions can tolerate higher $\dmse$ without significant increase of $\dssim$.
Therefore, many studies
used a content-complexity measure to normalize $\dmse$ as an approximation of $\dssim$ \cite{Dai2014ssim,qi2013efficient,yeo2013ssim, Yeo2013}.
They typically followed a
$\dssim$-$\dmse$ model proposed by Yeo \etal \cite{Yeo2013}:
\begin{equation}\label{equ:yeo}
  1/\text{SSIM}=1+{\dmse}/{(2\sigma^2+\epsilon)},
\end{equation}
where 1/SSIM is used as $\dssim$ in \cite{Yeo2013}, $\sigma^2$ is  variance of the block, and $\epsilon$ is a small constant.
The second column of \figref{fig:ddrelation} illustrates the modeling performance of \equref{equ:yeo}. Unfortunately, the $\dssim$-$\dmse$ modeling accuracy of this method is less than satisfactory for these videos.

In the studies of OBA (e.g. \cite{karczewicz2013intra,Gao2016}), sum of absolute transformed difference (SATD) is usually used to measure the content-complexity of a CTU.
It is the sum of the Hadamard transform coefficients of all the $L$ non-overlapped $8\times8$ sub-blocks in a CTU, which is calculated as
\begin{equation}\label{equ:satd}
S=\sum_{l=0}^L \sum_{m=0}^7 \sum_{n=0}^7 |h_{m,n,l}|-|h_{0,0,l}|,
\end{equation}
where $h_{m,n,l}$ indicates the Hadamard transformed coefficient at position $(m,n)$ in the $l$-th sub-block.

\begin{table}[]
\centering
\caption{PCCs of the relationships of $\dssim$-$\dmse$, variance-normalized $\dssim$-$\dmse$ \equref{equ:yeo}, and SATD-normalized $\dssim$-$\dmse$ (i.e., \equref{equ:globaldd}).}
\label{tab:pccs}
\begin{tabular}{ccccccc}
\toprule
& \multicolumn{2}{c}{$\textdssim$-$\textdmse$} &
\multicolumn{2}{c}{\equref{equ:yeo} \cite{Yeo2013}} & \multicolumn{2}{c}{$\textdssim$-$\textdmse/S$ } \\
\cmidrule(lr){2-3}
\cmidrule(lr){4-5}
\cmidrule(lr){6-7}
class & AI & LB & AI & LB & AI & LB \\
\midrule
A    & 0.20  & 0.23 & 0.18 & 0.22 & 0.81 & 0.70  \\
B    & 0.47 & 0.44 & 0.15 & 0.19 & 0.86 & 0.76 \\
C    & 0.40  & 0.31 & 0.32 & 0.31 & 0.86 & 0.66 \\
D    & 0.41 & 0.28 & 0.20 & 0.20 & 0.86 & 0.75 \\
E    & 0.52 & 0.61 & 0.11 & 0.11 & 0.76 & 0.77 \\
avg. & 0.42 & 0.38 & 0.19 & 0.20 & 0.84 & 0.73\\
\bottomrule
\end{tabular}
\end{table}

In the third column of \figref{fig:ddrelation}, we compare $\dssim$ with the SATD-normalized $\dmse$, indicated by $\dmse/S$.
We can see that $\dssim$ presents linear trend with $\dmse/S$.
\tabref{tab:pccs} reports the Pearson linear correlation coefficient (PCC) \cite{Taylor1990} between $\dssim$ and $\dmse/S$.
The PCCs between $\dssim$ and the $\sigma^2$-normalized $\dmse$ and PCCs between $\dssim$ and un-normalized $\dmse$ are also reported for comparison.
PCC lies in the range [-1,1]. The closer the PCC value is to 1/-1, the better the positive/negative linear correlation is.
The results in \tabref{tab:pccs} show that the SATD-based normalization greatly improves the linearity of $\dssim$-$\dmse$.
Therefore, a new SATD-normalization-based linear model is proposed to model the $\dssim$-$\dmse$ relationship in this study as follows:
\begin{equation}\label{equ:globaldd}
  {\dssim} = \Theta\cdot\dfrac{{\dmse}}{S}+H,
\end{equation}
where $\Theta$ and $H$ are the linear parameters.

By substituting \equref{equ:globaldd} into \equref{equ:costmap},
the SSIM-based RDO process can be re-written as
\begin{equation}\label{equ:costmapfinal}
  \begin{aligned}
  & \mathop{\arg\min}_{m}  {\dssim}_{i,j} (m) \cdot\text{M}_{i,j}+{\lambda^*_\text{SSIM}}_{i,j} \cdot R_{i,j}(m)\\
  =& \mathop{\arg\min}_{m}  (\Theta\cdot\dfrac{{\dmse}_{i,j}(m)}{S_{i,j}}+H) \cdot\text{M}_{i,j}+{\lambda^*_\text{SSIM}}_{i,j} \cdot R_{i,j}(m)\\
  =& \mathop{\arg\min}_{m} {\dmse}_{i,j}(m) + \frac{S_{i,j}}{\Theta} {\lambda^*_\text{SSIM}}_{i,j} \cdot R_{i,j}(m)\\
  \end{aligned}
\end{equation}
According to \equref{equ:costmapfinal},
the SSIM-based RDO can be achieved based on the R-$\dmse$ cost with a new Lagrangian multiplier denoted by $\lambda_\text{MSE}^{new}$ that is a scaled $\lssim$, i.e.,
\begin{equation}\label{equ:ll}
 {\lambda_\text{MSE}^{new}}_{i,j} = \frac{S_{i,j}}{\Theta}{\lambda^*_\text{SSIM}}_{i,j}.
\end{equation}
In this way, the $\dssim$ does not need to be calculated for all candidate modes in the SSIM-based RDO process, thus saving the encoding time.
Besides,
${\lambda^*_\text{SSIM}}_{i,j}$ has been calculated by the SSIM-based OBA process \equref{equ:optimallambda}.
Thus, only $S$  and $\Theta$ need to be calculated to obtain the
${\lambda_\text{MSE}^{new}}_{i,j}$.
Estimation of $\Theta$ for each CTU will be proposed in \secref{subsec:ddpestimation}.

After encoding $\text{CTU}_{i,j}$ with ${\lambda}^{new}_{\text{MSE}_{i,j}}$, the ${\dssim}_{i,j}$ and $bpp_{i,j}$ are generated.
Then, the joint R-D-$\lambda$ relationship-based parameter estimation strategy
can be used to calculate the unknown model parameters.
Specifically, in the joint R-D-$\lambda$ relationship composed of \equref{equ:rdmodel} and \equref{equ:rlambda}:
\begin{equation}\label{equ:joint}
  \left\{
    \begin{aligned}
        {\dssim}_{i,j} &= \alpha_{i,j} \cdot {bpp}_{i,j}^{\beta_{i,j}},\\
        {\lssim}_{i,j}  &= -\alpha_{i,j} \beta_{i,j} \cdot {bpp}_{i,j}^{\beta_{i,j} -1},
    \end{aligned}
  \right.
\end{equation}
only $\alpha_{i,j}$ and $\beta_{i,j}$ are unknown.
Therefore, they can be uniquely solved, and the solved parameters will be used to solve the SSIM-based OBA problem of the subsequent frame, which can be described as:
\begin{equation}\label{equ:ckssim}
\left\{
\begin{aligned}
&\alpha_{i,j+1}\approx\alpha_{i,j} =\dfrac{{\dssim}_{i,j}} { {{bpp}_{i,j}}^{-\frac{\Theta_{i,j}}{S_{i,j}} {\lambda}^{new}_{\text{MSE}_{i,j}}  {bpp}_{i,j}/{\dssim}_{i,j}}},\\
&\beta_{i,j+1}\approx\beta_{i,j} = -\dfrac{\frac{\Theta_{i,j}}{S_{i,j}} {\lambda}^{new}_{\text{MSE}_{i,j}} {bpp}_{i,j} }{{\dssim}_{i,j}}.
\end{aligned}
\right.
\end{equation}

Then, combined with the SSIM-based OBA scheme that has been described in SOMR, OBA and RDO can be jointly optimized based on SSIM.
Besides, it is worth noting that although the R-D-$\lambda$ joint relationship has been used in many MSE-based works such as \cite{Li2017}, this is the first time it is exploited in the SSIM-based studies.
Moreover, without the $\lssim$-$\lmse$ relationship \equref{equ:ll} proposed in this study,
it is unknown which $\lssim$ is associated with the R and $\dssim$ generated by encoding, then the R-$\dssim$-$\lssim$ joint relationship cannot be solved.
Therefore, the proposed method is of great significance not only for optimizing OBA and RDO jointly based on SSIM, but also for improving the estimation accuracy of R-$\dssim$ model parameters.

\subsection{Parameter Estimation of $\dssim$-$\dmse$ model}\label{subsec:ddpestimation}
In this subsection, estimation of the $\dssim$-$\dmse$ model parameters is proposed.
Since the proposed $\dssim$-$\dmse$ model is a linear model, the method of linear regression can be applied for its parameter estimation.
However, as mentioned in the introduction section,
there is a limitation of the regression method for R-D modeling, that is, it is invalid when the contents of a series of CTUs change.
Unlike the traditional hyperbolic or exponential R-D models,
the proposed $\dssim$-$\dmse$ model extracts content-related SATD for each CTU, which brings adaptability to the change of content and therefore partially solves the limitation of the regression method.
Moreover,
because the collocated CTUs in adjacent frames usually have similar content, this can be exploited to further weaken the impact of content changes on $\dssim$-$\dmse$ modeling.
In the right part of \figref{fig:ddrelation}, we plot $\dssim$-$\dmse/S$ for several randomly selected examples of the adjacently collocated CTUs.
It can be seen that most of them present good linearity.
Based on this observation, we apply the proposed linear $\dssim$-$\dmse$ model for each CTU in a frame separately.
In other words, each CTU in a frame will possess different $\dssim$-$\dmse$ model parameters, and these parameters are estimated based on the true $\dssim$-$\dmse$ values of the respective collocated CTUs.
Specifically,
for $\text{CTU}_{i,j}$, the proposed $\dssim$-$\dmse$ model can be described as:
\begin{equation}\label{equ:lineardd}
  {\dssim}_{i,j} = \theta_{i,j} \cdot {\dfrac{{\dmse}_{i,j}}{S_{i,j}}} + \eta_{i,j},
\end{equation}
where $\theta_{i,j}$ and $\eta_{i,j}$ are the linear model parameters.
After $\text{CTU}_{i,j}$ is encoded, the resulting actual ${\dssim}_{i,j}$ and ${\dmse}_{i,j}$ will be used to estimate the model parameters in the adaptive least mean square method \cite{widrow1960adaptive} as follows:
\begin{equation}\label{equ:slope}
  \left\{
  \begin{aligned}
  &\Delta D={\dssim}_{i,j}\!-\!\theta_{i,j} \cdot \dfrac{{\dmse}_{i,j}}{S_{i,j}}\!-\!\eta_{i,j}\\
  & \theta_{i,j+1}\approx\theta_{i,j} = \theta_{i,j} + \delta_\theta \cdot \Delta D \cdot {{\dmse}_{i,j}},\\
  & \eta_{i,j+1}\approx\eta_{i,j}=\eta_{i,j} + \delta_\eta \cdot \Delta D,
  \end{aligned}
  \right.
\end{equation}
where $\delta_\theta$ and $\delta_\eta$ control the updating speed, both of which are empirically set as $0.01$; $\theta_{i,j+1}$ and $\eta_{i,j+1}$ will be used for the collocated CTU in subsequent frame. The initial values of $\theta$ and $\eta$ are empirically set as $S_{i,0}\cdot{\dssim}_{i,0}/{\dmse}_{i,0}$ and $0$, respectively.

To evaluate the modeling performance,
we calculate the prediction error between the actual and the predicted values of $\dssim$ with different models as follows:
\begin{equation}\label{equ:error}
    P_{i,j}={\left|{\dssim}_{i,j}-g({\dmse}_{i,j})\right|}/{{\dssim}_{i,j}},
\end{equation}
where $g$ indicates the $\dssim$-$\dmse$ model, and ${\dssim}_{i,j}$ and ${\dmse}_{i,j}$ are the actual encoding results.
Prediction error of three models including Yeo's model \cite{Yeo2013} (i.e.,\equref{equ:yeo}), the proposed model directly applied for all the CTUs (i.e., \equref{equ:globaldd}), and the proposed model applied for each the collocated CTUs (i.e., \equref{equ:lineardd}) are denoted by $P_{yeo}$, $P_{global}$, and $P_{local}$, respectively.

\begin{table}[t]
\caption{Average prediction error of three $\dssim$-$\dmse$ models over all CTUs of videos.}
\label{tab:error}
\centering
\begin{tabular}{ccccccc}
\toprule
      & \multicolumn{3}{c}{AI}               & \multicolumn{3}{c}{LB}               \\
      \cmidrule(lr){2-4}
      \cmidrule(lr){5-7}
class & $P_\text{yeo}$ & $P_\text{global}$ & $P_\text{local}$ & $P_\text{yeo}$ & $P_\text{global}$ & $P_\text{local}$ \\
\midrule
A     & 95.3\% & 17.8\%           & 6.4\%    &95.4\%       & 20.5\%           & 17.5\%          \\
B     & 93.8\% & 17.8\%           & 6.6\%    &93.9\%       & 21.3\%           & 12.9\%          \\
C     &96.0\% & 17.6\%           & 9.6\%     &96.1\%      & 23.6\%           & 19.6\%          \\
D     &96.2\% & 17.4\%           & 8.2\%     &96.1\%      & 20.7\%           & 16.5\%          \\
E     &96.7\% & 27.3\%           & 5.0\%     &96.7\%      & 28.4\%           & 7.8\%           \\
avg.  &95.5\% & 19.3\%           & 7.3\%     &95.5\%      & 22.8\%           & 14.8\% \\
\bottomrule
\end{tabular}
\end{table}

\tabref{tab:error} lists the average $P_{i,j}$ over all CTUs of videos that are encoded in AI and LB configurations at four QPs.
Results show that
prediction error of the proposed model \equref{equ:lineardd} (i.e., $P_{local}$) is much smaller than that of the other two models in both AI and LB configurations.
Even for in the LB configuration where the parameter updating \equref{equ:slope} is performed between two adjacent frames at the same hierarchical level, which can be separated by a maximum of four frames in temporal, $P_{local}$ still has a better performance than the other two models.

\begin{algorithm}[t]
\caption{SOSR: unified optimization of SSIM-based OBA and SSIM-based RDO}
\begin{algorithmic}
   \Require{$\alpha_{i,j}$, $\beta_{i,j}$, $\theta_{i}$, $i\in\{0,1,\cdots N\}$}
       \Procedure{1: \textit{SSIM-based Optimal Bit Allocation}}{}
       \State  Search the optimal ${\lambda^*_\text{SSIM}}_{i,j}$ and calculate the corresponding $R^*_{i,j}$ that satisfies \equref{equ:optimallambda} with the Bisection method;
       \EndProcedure
       \Procedure{2: \textit{CTU Encoding With SSIM-Based RDO}}{}
       \For{$i=0:N-1$}
       \State a. Calculate SATD $S_{i,j}$ of the CTU;
       \State b. SSIM-Based Bitrate Control:
       \State\hskip2.5em  Calculate ${\lambda_\text{MSE}^{new}}_{i,j}$ by \equref{equ:ll};
       \State c. Traverse all possible encoding modes;
       \State d. {{\hangafter0\hangindent1em} Search the mode with minimum R-$\dmse$ cost with the new Lagrangian multiplier ${\lambda_\text{MSE}^{new}}_{i,j}$;}
       \State e. Encode  $\text{CTU}_{i,j}$ with the selected mode;
       \EndFor
       \EndProcedure
       \Procedure{3: \textit{Parameter Estimation For Subsequent Frame}}{}
       \State Calculate the SSIM map of encoded frame by \equref{equ:ssimmap};
       \For{$i=0:N-1$}
       \State a. Record $bpp_{i,j}$, ${\dssim}_{i,j}$ and, ${\lambda^{new}_\text{MSE}}_{i,j}$;
       \State b. Update the linear model parameter by \equref{equ:slope};
       \State c. Estimate $\alpha_{i,j+1}$ and $\beta_{i,j+1}$ by \equref{equ:ckssim};
       \State d. Estimate $\theta_{i,j+1}$ and $\beta_{i,j+1}$ by \equref{equ:slope};
       \EndFor
       \EndProcedure
\end{algorithmic}
\end{algorithm}

\section{Experiments}
In this section, experimental comparison is conducted to demonstrate advantage of the proposed SOSR scheme.
First,
the R-D performance is evaluated to verify the effectiveness of the proposed scheme.
Besides, comprehensive analyses of the proposed scheme are presented.

\subsection{Experiment Setup}
We have implemented the proposed SOSR into HM16.20.
The whole process to encode one frame is summarized in detail in Algorithm~1.

For performance comparison,
because many SSIM-based RDO studies did not study SSIM-based OBA, the encoding bitrate constraint usually cannot be accurately achieved \cite{Wang2012ssim,wang2013perceptual,Yeo2013,Dai2014ssim,qi2013efficient,yeo2013ssim}.
Moreover, many of them are specified for the previous H.264 standard
\cite{Wang2012ssim,wang2013perceptual,Yeo2013,Dai2014ssim}.
Thus, for fair comparison,
six state-of-the-art OBA schemes for HEVC are adopted as competitors, including JCTVC-K0103 \cite{li2012rate}, JCTVC-M0257 \cite{karczewicz2013intra}, and Gao \cite{Gao2016} for intra encoding; JCTVC-M0036 \cite{li2013adaptive}, Li \cite{Li2017} and Zhou \cite{Zhou2019} for inter encoding.
In these schemes, Gao \cite{Gao2016} and Zhou \cite{Zhou2019} are respectively the SSIM-based OBA schemes for intra and inter encoding, while the others are based on MSE.

For performance evaluation,
we use the same setup as \cite{Gao2016} and \cite{Zhou2019}.
Specifically,
AI and LDB (h and nh) configurations are adopted.
In each configuration, the standard test video sequences from class A to class E are encoded at four QPs (22, 27, 32, and 37) \cite{bossen2013common}.
Then, the resulted bitrates are set as the bit constraints for encoding of an OBA scheme.
For comparison,
JCTVC-K0103 and JCTVC-M0036 are set as the anchor schemes for intra and inter encoding, respectively.
Specifically,
by comparing a scheme to the anchor, the R-$\dssim$ performance of the scheme is evaluated in terms of BDBR and BD-SSIM \cite{bdrate}.
Besides, it is worth noting that different implementations of SSIM will yield different BD-SSIM values.
Thus, BDBR which is a relative measure is more credible to verify the R-$\dssim$ performance than BD-SSIM.
In this paper, the standard implementation of SSIM \cite{Wang2004} that is available at \cite{ssimweb} is adopted.

\begin{table}[t]
\centering
\caption{R-D performance comparison for intra encoding in terms of BDBR and BR-SSIM.
Configuration: AI.}
\label{tab:intrabdresults}
\begin{tabular}{*{7}c}
\toprule
      & \multicolumn{4}{c}{anchor: K0103\cite{li2012rate}} & \multicolumn{2}{c}{anchor: M0257\cite{karczewicz2013intra}}\\
      \cmidrule(lr){2-5}
      \cmidrule(lr){6-7}
      & \multicolumn{2}{c}{BDBR} & \multicolumn{2}{c}{BD-SSIM} & BDBR & BD-SSIM \\
\cmidrule(lr){2-3}
\cmidrule(lr){4-5}
\cmidrule(lr){6-7}
class & Gao\cite{Gao2016}        & Ours        & Gao\cite{Gao2016}    & Ours     &Ours &Ours    \\
\midrule
A     & -2.0       & -14.4       & 0.0010        & 0.0008   & -6.6  & 0.0003\\
B     & -2.0       & -9.9        & 0.0009       & 0.0012  & -5.0  & 0.0006\\
C     & -4.1       & -8.2        & 0.0023       & 0.0031  & -3.2  & 0.0012\\
D     & -4.1       & -3.2        & 0.0022       & 0.0029  & -1.0  & 0.0009 \\
E     & -1.6       & -8.9        & 0.0003       & 0.0010    & -6.6  & 0.0007\\
avg.  & -2.7       & -8.9        & 0.0013       & 0.0018  & -4.5  & 0.0007\\
\bottomrule
\end{tabular}
\end{table}

\begin{table}[t]
\caption{R-D performance comparison for inter encoding in terms of BDBR and BD-SSIM.
Anchor: JCTVC-M0036 \cite{li2013adaptive}.
Configuration: h\_LB.
}
\label{tab:interbdresults_h}
\centering
{
\begin{tabular}{*{7}c}
\toprule
     & \multicolumn{3}{c}{{BDBR}}          & \multicolumn{3}{c}{BD-SSIM} \\
\cmidrule(lr){2-4}
\cmidrule(lr){5-7}
class & Li\cite{Li2017} & Zhou\cite{Zhou2019} & Ours    & Li \cite{Li2017} & Zhou\cite{Zhou2019} & Ours  \\
\midrule
A     & -5.4       & -5.8        & -17.0   & 0.0003         & 0.0030      & 0.0008 \\
B     & -4.9      & -4.9        & -13.1    & 0.0006         & 0.0027      & 0.0010 \\
C     & -1.8      & -4.9        & -7.3     & 0.0006          & 0.0027      & 0.0022 \\
D     & -1.6      & -12.2       & -8.2     & 0.0011          & 0.0074      & 0.0048 \\
E     & -4.6      & -5.1        & -8.5      & 0.0004          & 0.0027      & 0.0006 \\
avg.  & -3.7     & -6.6         &-10.8    & 0.0006         & 0.0037      & 0.0019 \\
\bottomrule
\end{tabular}
}
\end{table}

\begin{table}[t]
\caption{R-D performance comparison for inter encoding in terms of BDBR and BD-SSIM. Anchor: JCTVC-M0036 \cite{li2013adaptive}.
Configuration: nh\_LB.
}
\label{tab:interbdresults_nh}
\centering
{
\begin{tabular}{*{7}c}
\toprule
     & \multicolumn{3}{c}{{BDBR}}          & \multicolumn{3}{c}{BD-SSIM} \\
\cmidrule(lr){2-4}
\cmidrule(lr){5-7}
class & Li\cite{Li2017} & Zhou\cite{Zhou2019} & Ours    & Li \cite{Li2017} & Zhou\cite{Zhou2019} & Ours   \\
\midrule
A     & -8.4      & -11.7       & -22.2   & 0.0005  & 0.0067      & 0.0013 \\
B     & -9.8      & -13.9       & -16.5   & 0.0016  & 0.0076      & 0.0028 \\
C     & -3.1      & -12.3       & -9.2    & 0.0012  & 0.0074      & 0.0039 \\
D     & -3.0      & -22.8       & -12.8   & 0.0022  & 0.0139      & 0.0093 \\
E     & -5.2      & -9.4        & -12.0    & 0.0005    & 0.0064    & 0.0011 \\
avg.  & -5.9     & -14.0      & -14.5   & 0.0012   & 0.0084      & 0.0037  \\
\bottomrule
\end{tabular}
}
\end{table}

\subsection{R-D Performance Comparison}

For intra encoding, \tabref{tab:intrabdresults} summarizes the performance comparison results.
Compared with JCTVC-K0103, the proposed SOSR scheme achieves 8.9\% BDBR saving and 0.0018 BD-SSIM improvement, both of which are better than the results of Gao's scheme.
In addition to JCTVC-K0103, the proposed scheme is also compared with JCTVC-M0257 that is the default intra OBA scheme of HM16.20.
As shown in \tabref{tab:intrabdresults}, the BDBR and BD-SSIM gains of the proposed model are 4.5\% and 0.0007, respectively, which further verify the effectiveness of the proposed model.

For inter encoding, both h\_LB and nh\_LB configurations are adopted in the comparison.
The results are respectively summarized in \tabref{tab:interbdresults_h} and \tabref{tab:interbdresults_nh}, where JCTVC-M0036 is
set as the anchor.
Li's scheme is implemented in HM16.20 based on their source codes.
The results of Zhou's scheme are from their paper \cite{Zhou2019}.
It can be seen from the results that
comparing with Li's scheme, the proposed SOSR scheme presents much better performance in terms of both BDBR and BD-SSIM.
Besides, Zhou's scheme achieves larger BD-SSIM than the proposed scheme.
This might be due to the different implementation of SSIM.
We can see that the average BDBR saving of our scheme are 10.8\% and 14.5\% in h\_LB and nh\_LB configurations, respectively, both of which are better than that of Zhou's scheme.
As has been discussed, BDBR measures the relative increase of bitrate, while BD-SSIM measures the absolute gain of SSIM.
When different implementations of SSIM are used, the larger BDBR saving is more credible to verify the better R-$\dssim$ performance of the proposed SOSR scheme.

\subsection{Analysis of the Proposed Model}\label{sec:analysis}

\begin{table*}[]
\caption{Average bits error at CTU level. Symbol `---' indicates that the result was not provided in the corresponding paper.}
\label{tab:biterror}
\centering
\begin{tabular}{*{15}c}
\toprule
& \multicolumn{4}{c}{AI}    & \multicolumn{5}{c}{h\_LB}        & \multicolumn{5}{c}{nh\_LB}       \\
    \cmidrule(lr){2-5}
    \cmidrule(lr){6-10}
    \cmidrule(lr){11-15}
class & \rotatebox{60}{M0257} & \rotatebox{60}{SOMR} & \rotatebox{60}{SOMR+} & \rotatebox{60}{SOSR} & \rotatebox{60}{M0036}   & \rotatebox{60}{Li\cite{Li2017}}  & \rotatebox{60}{SOMR} & \rotatebox{60}{SOMR+} & \rotatebox{60}{SOSR} & \rotatebox{60}{M0036}    & \rotatebox{60}{Li\cite{Li2017}}  & \rotatebox{60}{SOMR} & \rotatebox{60}{SOMR+} & \rotatebox{60}{SOSR} \\
\midrule
a          & 0.3 & 124.2 & 0.2 & 0.1 & 7.3  & --  & 28.7 & 12.6 & 2.8 & 5.4  & --  & 36.1 & 5.3  & 1.2\\
b          & 1.9 & 45.8  & 6.1 & 0.4 & 10.9 & 1.4 & 28.3 & 12.6 & 2.4 & 9.3  & 4.5 & 27.9 & 6.7  & 1.1 \\
c          & 2.6 & 22.2  & 1.7 & 1.1 & 22.7 & 1.2 & 51.1 & 18.4 & 3.2 & 13.6 & 4.3 & 27.7 & 7.7  & 1.6 \\
d          & 3.2 & 18.2  & 1.6 & 2.1 & 11.0 & 0.9 & 48.5 & 25.4 & 4.7 & 7.2  & 2.1 & 25.5 & 12.8 & 3.5 \\
e          & 0.8 & 41.4  & 0.4 & 0.2 & 2.9  & 1.1 & 12.1 & 8.1  & 1.4 & 4.2  & 1.9 & 14.1 & 9.2  & 0.9 \\
avg. (b-e) & 2.2 & 32.2  & 2.8 & 1.0 & 12.4 & 1.2 & 36.0 & 16.4 & 3.0 & 8.9  & 3.4 & 24.7 & 8.9  & 1.8 \\
\bottomrule
\end{tabular}
\end{table*}

\begin{table}[]
\caption{Comparison with SOMR and SOMR+ in terms of BDBR.
Anchor: M0257 for AI, M0036 for h\_LB and nh\_LB.
}
\label{tab:comparewithreg}
\centering
\resizebox{\linewidth}{!}
{
\begin{tabular}{cccccccccc}
\toprule
    & \multicolumn{3}{c}{AI} & \multicolumn{3}{c}{h\_LB} & \multicolumn{3}{c}{nh\_LB} \\
    \cmidrule(lr){2-4}
    \cmidrule(lr){5-7}
    \cmidrule(lr){8-10}
class & \rotatebox{60}{SOMR}   & \rotatebox{60}{SOMR+} & \rotatebox{60}{SOSR}  & \rotatebox{60}{SOMR}   & \rotatebox{60}{SOMR+} & \rotatebox{60}{SOSR}   & \rotatebox{60}{SOMR}    & \rotatebox{60}{SOMR+} & \rotatebox{60}{SOSR}   \\
\midrule
A     & -1.9 & -3.1  & -6.6 & -7.6 & -7.8  & -17.0 & -11.1 & -13.5  & -22.2 \\
B     & -0.1 & -3.8  & -5.0 & -5.8 & -2.7  & -13.1 & -10.1 & -10.3  & -16.5 \\
C     & 1.1  & 0.9  & -3.2 & -2.0  & -1.6  & -7.3  & -3    & -4.1  & -9.2 \\
D     & -0.1 & 0.1  & -1.0 & -3.4 & -4.5  & -8.2  & -4.3  & -5.4  & -12.8 \\
E     & -2.3 & -4.0  & -6.6 &  0.7  & -2.3  & -8.5  & -0.5  & -5.2  & -12.0  \\
avg.  & -0.7 & -2.0  & -4.5 & -3.5 & -3.8  & -10.8& -5.7  & -7.7  & -14.5\\
\bottomrule
\end{tabular}
}
\end{table}
In \tabref{tab:comparewithreg}, we summarized the R-$\dssim$ performance of SOMR and SOSR in terms of BDBR.
The results verify that the proposed SOSR has much better
R-$\dssim$ performance than the conventional SOSR.
Recall that compared with the conventional SOMR scheme,
the proposed SOSR scheme has two advantages: the joint R-$\dssim$-$\lssim$-based parameter estimation and the unified optimization of SSIM-based OBA and SSIM-based RDO.
To further analyze the respective contributions of these two advantages,
we replaced calculation of the R-$\dssim$ parameters in SOMR with the R-$\dssim$-$\lssim$ joint relationship-based estimation.
The resulted scheme is denoted by SOMR+.
Specifically, in SOMR+, the $\lmse$ for RDO is still calculated based on the default bitrate control scheme of HM that is based on MSE, i.e., \equref{equ:add_rl}.
After encoding the CTU, $\lmse$ can be mapped to $\lssim$ according to the proposed $\lssim$-$\lmse$ relationship \equref{equ:ll}.
The R-$\dssim$ model parameters are then calculated based on the mapped $\lssim$ and the actual R and $\dssim$ of the CTU.
That is,
on the basis of SOMR, SOMR+ adds the joint R-$\dssim$-$\lssim$ relationship-based parameter estimation, and SOSR optimizes SSIM-based OBA and SSIM-based RDO jointly further.

The BDBR of SOMR+ are also listed in \tabref{tab:comparewithreg}.
Compared with SOMR,
SOMR+ respectively achieves 1.3\%, 0.3\%, and 2\% BDBR savings in the three configurations,
while SOSR further saves 2.5\%, 7\%, and 6.8\% BDBR compared with SOMR+.
The results show that the R-$\dssim$ performance can be improved with more accurate R-$\dssim$ model.
At the same time, with unified optimization of SSIM-based OBA and SSIM-based RDO in the proposed SOSR, the R-$\dssim$ performance can be improved further.
Therefore, we can conclude that
the two advantages of SOSR together result in its outstanding R-$\dssim$ performance.

In addition to the R-$\dssim$ performance,
we evaluate the bits error at CTU level of different schemes, which is the difference between the allocated bits $R^*_{i,j}$ and the actual encoding bits $R^{act}_{i,j}$ of a CTU:
\begin{equation}\label{equ:biterror}
E_{i,j} = {\left|R^*_{i,j}-R^{act}_{i,j}\right|}/{R^*_{i,j}}.
\end{equation}
The average values of $E_{i,j}$ of different schemes are summarized in \tabref{tab:biterror}.
Since encoding bits are controlled by $\lmse$,
the smaller bits error demonstrates the more accurate R-$\lmse$ modeling performance.
As \tabref{tab:biterror} shows,
JCTVC-M0257 and JCTVC-M0036 that use regression method to estimate the R-$\lmse$ model parameters have large bits error.
The bits error is further expanded in SOMR that uses regression method for both R-$\dssim$ and R-$\lmse$ estimation.
Besides, for \cite{Li2017}, the bits error has been significantly reduced, which is due to its joint R-$\dmse$-$\lmse$ based parameter estimation that improves accuracy of the R-$\lmse$ model.
For SOMR+,
using the proposed $\lssim$-$\lmse$ relationship, more accurate R-$\dssim$ model can be achieved with based on the joint R-$\dssim$-$\lssim$ relationship.
However, without unified optimization of SSIM-based OBA and SSIM-based RDO, accuracy of the R-$\lmse$ model will not be improved.
On the other hand, this problem is solved in SOSR.
Results show that the bits error of SOSR is also significantly reduced.

\subsection{SSIM vs. PSNR}
\begin{table}[]
\centering
\caption{R-D performance comparison for intra encoding in terms of BDBRp and BD-PSNR. Symbol `---' indicates that the result was not provided in the corresponding paper.
Configuration: AI.
}
\label{tab:intrabdresults_psnr}
\begin{tabular}{*{7}c}
\toprule
      & \multicolumn{4}{c}{anchor: K0103\cite{li2012rate}} & \multicolumn{2}{c}{anchor: M0257\cite{karczewicz2013intra}}\\
      \cmidrule(lr){2-5}
      \cmidrule(lr){6-7}
      & \multicolumn{2}{c}{BDBRp} & \multicolumn{2}{c}{BD-PSNR} & BDBRp & BD-PSNR \\
    \cmidrule(lr){2-3}
    \cmidrule(lr){4-5}
    \cmidrule(lr){6-7}
class & Gao\cite{Gao2016}        & Ours        & Gao\cite{Gao2016}          & Ours & Ours & Ours       \\
\midrule
A     & ---           & 3.05        & 0.08         & -0.17   &2.18 &-0.13    \\
B     & ---           & 4.37        & 0.06         & -0.17   &1.96 &-0.08    \\
C     & ---           & 2.93        & 0.15         & -0.16   &1.93 &-0.11    \\
D     & ---           & 2.44        & 0.19         & -0.17   &1.04 &-0.07    \\
E     & ---           & 4.78        & 0.07         & -0.24   &2.24 &-0.12    \\
avg. & ---           & 3.51        & 0.11         & -0.18    &1.87 &-0.10  \\
\bottomrule
\end{tabular}

\end{table}
\begin{table}[]
\caption{R-D performance comparison for inter encoding in terms of BDBRp and BD-PSNR.
Anchor: JCTVC-M0036 \cite{li2013adaptive}.
Configuration: h\_LB.
}
\label{tab:interbdresults_h_psnr}
\centering
\resizebox{\linewidth}{!}
{
\begin{tabular}{*{7}c}
\toprule
     & \multicolumn{3}{c}{BDBRp}        & \multicolumn{3}{c}{BD-PSNR}       \\
\cmidrule(lr){2-4}
\cmidrule(lr){5-7}
class & Li\cite{Li2017} & Zhou\cite{Zhou2019} & SOSR    & Li \cite{Li2017} & Zhou\cite{Zhou2019} & SOSR  \\
\midrule
A     & -2.1     & -3.2        & -0.9   & 0.07         & 0.12        & -0.01 \\
B     & -2.6     & -3.0        & 2.7    & 0.06      & 0.11        & -0.06   \\
C     & -0.8     & -2.7        & 1.4    & 0.03      & 0.09        & -0.06   \\
D     & -0.7     & -3.3        & 1.2    & 0.03      & 0.13        & -0.06   \\
E     & -5.8     & -3.2        & 1.5    & 0.17       & 0.12        & -0.02   \\
avg. & -2.4     & -3.1        & 1.2    & 0.07         & 0.11        & -0.04\\
\bottomrule
\end{tabular}

}
\end{table}

\begin{table}[]
\caption{R-D performance comparison for inter encoding in terms of BDBRp and BD-PSNR.
Anchor: JCTVC-M0036 \cite{li2013adaptive}.
Configuration: nh\_LB.
}
\label{tab:interbdresults_nh_psnr}
\centering
\resizebox{\linewidth}{!}
{
\begin{tabular}{*{7}c}
\toprule
       & \multicolumn{3}{c}{BDBRp}        & \multicolumn{3}{c}{BD-PSNR}       \\
\cmidrule(lr){2-4}
\cmidrule(lr){5-7}
class & Li\cite{Li2017} & Zhou\cite{Zhou2019} & SOSR    & Li \cite{Li2017} & Zhou\cite{Zhou2019} & SOSR  \\
\midrule
A     & -2.6      & -5.2        & -0.1 & 0.08       & 0.23        & -0.07 \\
B     & -4.9      & -2.9        & 8.4   & 0.12      & 0.1         & -0.17 \\
C     & -2.2      & -5.5        & 5.7    & 0.09      & 0.25        & -0.19 \\
D     & -1.8      & -5.8        & 2.0   & 0.07      & 0.26        & -0.08 \\
E     & -4.3      & -5.5        & 1.5    & 0.13      & 0.25        & -0.02 \\
avg.  & -3.2     & -5.0        & 3.5    & 0.10      & 0.22        & -0.11\\
\bottomrule
\end{tabular}
}
\end{table}

In addition to SSIM, the MSE-based peak signal-to-noise ratio (PSNR) is also widely used to evaluate quality of an encoded video.
Therefore, in this subsection, we calculate
the PSNR-based BDBR (denoted by BDBRp) and BD-PSNR of different schemes for reference.
The results are shown in \tabref{tab:intrabdresults_psnr} to \tabref{tab:interbdresults_nh_psnr} for AI, h\_LB and nh\_LB, respectively.
As shown in these tables, the proposed SOSR is inferior to other schemes in terms of BDBRp and BD-PSNR.
Compared with HM16.20, our scheme has 1.87\%, 1.2\%, 3.5\% bit increase under the same PSNR in AI, h\_LB, and nh\_LB configurations, respectively.
This is not surprising, because PSNR is calculated based on MSE, which is not the optimization objective of our scheme.

\begin{figure*}[]
    \centering
    \includegraphics[width=\linewidth]{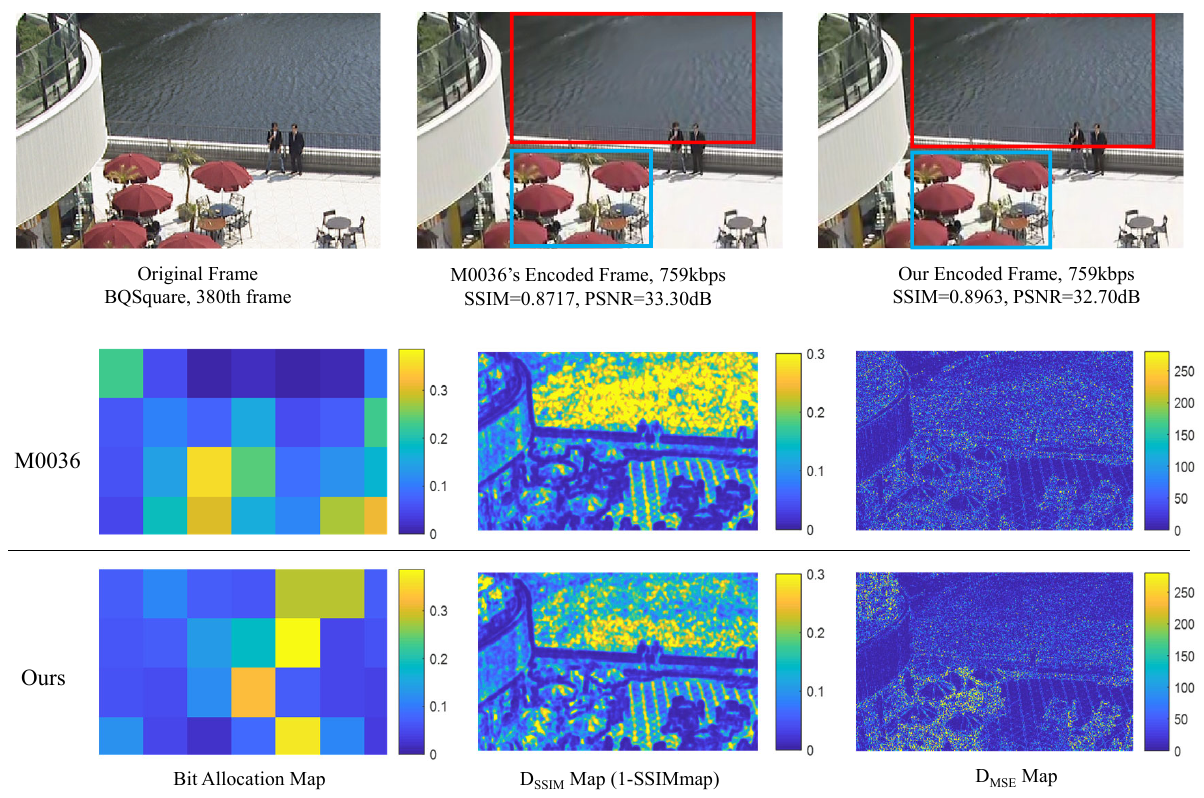}
    \caption{Visual comparison of frames encoded by JCTVC-M0036 \cite{li2013adaptive} and the proposed SOSR. The $\dssim$ map is calculated by 1-SSIMmap. The $\dmse$ map illustrates the pixel-wise $\dmse$. Example frame: BQSquare, 380th frame, 759kbps, nh\_LB.
    } 
    \label{fig:frame}
\end{figure*}

\begin{figure*}[]
    \centering
    \includegraphics[width=\linewidth]{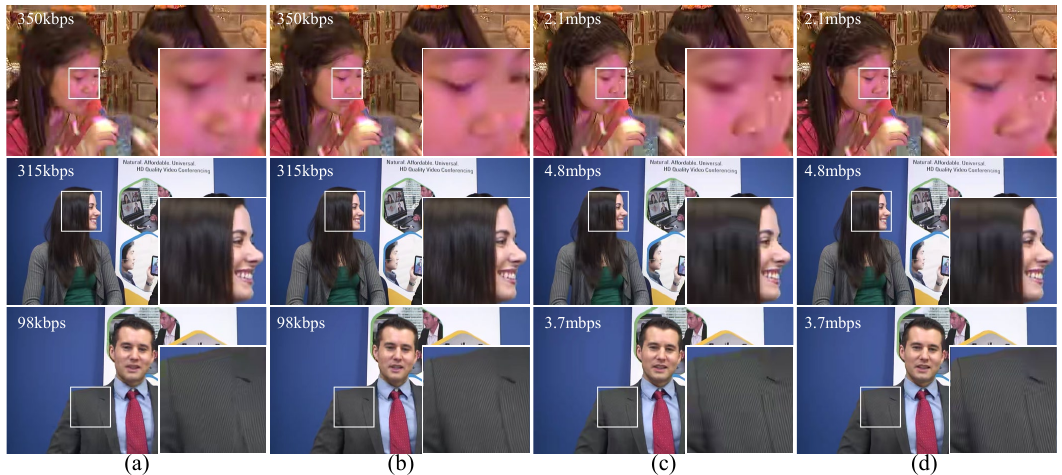}
    \caption{Visual quality comparison of frames encoded by the proposed SOSR scheme, Gao \cite{Gao2016}, and Li \cite{Li2017} at different bitrate.
    The frames from the first to the last row are BlowingBubbles 48th frame, KristenAndSara 60th frame, and Johnny 60th frame. (a) Li \cite{Li2017}, nh\_LB. (b) SOSR, nh\_LB. (c) Gao \cite{Gao2016}, AI. (d) SOSR, AI.
    } 
    \label{fig:compare}
\end{figure*}

\figref{fig:frame} illustrates
the difference in encoding results caused by different optimization objectives, where an example frame is respectively encoded by our scheme and JCTVC-M0036 at same bitrate.
JCTVC-M0036 minimizes $\dmse$ of a frame and is the default inter OBA scheme of HM16.20.
Therefore, the encoded frame has a better PSNR than the frame encoded by our scheme.
However, we can find that the frame encoded by \cite{li2013adaptive} does not have a good visual quality compared with that encoded by our scheme.

Specifically, as shown in the bit allocation map, JCTVC-M0036 allocates a large number of bits to the region marked in blue box and a small number of bits to region marked in red box.
Among them, the blue boxed region is highly textured, while the red boxed region has simpler structures.
We can see that after encoding with JCTVC-M0036, all these regions do not have significant $\dmse$ distortion.
However, the water ripple, i.e., the red boxed region with lost a lot of structural information, has become very smooth after encoding.
These structural distortions will attract the attention of the viewer and lead to the decline of subjective perceived quality.

On the other hand, we can see that $\dssim$ map correctly identifies these structural distortions.
Correspondingly, our scheme allocates more bits to the water ripple region.
It can be clearly seen that structure of this region is preserved and the visual quality is improved.
At the same time, because the total bits are constrained,
the bits allocated to the blue boxed region by our scheme are reduced, so this region has larger $\dmse$ distortion.
However, because of the visual masking effect \cite{ross1991contrast},
quality of the blue boxed region is good, whether based on the visual observation or the $\dssim$ map.
Thus, the frame encoded by our scheme has better visual quality. Therefore, it is reasonable to use $\dssim$ as the distortion metric in the optimization objective.

In \figref{fig:compare},
we further compare the subjective quality performance of the proposed SOSR with two available encoders Gao \cite{Gao2016} and Li \cite{Li2017}.
Both Gao \cite{Gao2016} and Li \cite{Li2017} achieves better R-$\dssim$ performance than the default HM.
\figref{fig:compare} shows that the frames encoded proposed SOSR achieves better visual quality at a similar bitrate than the other two schemes.
In particular, SOSR can better retain structural information in different texture areas, while other schemes bring more severe distortions, such as blurring, which reduces visual quality.

\begin{table}[t]
\centering
\caption{Encoding time comparison.}\label{tab:timecomparison}
\begin{tabular}{llllll}
\toprule
  &Gao\cite{Gao2016} &Li \cite{Li2017}       &Zhou\cite{Zhou2019} & SOSR  \\
\midrule
AI     &101.0\%     &-        &-        & 100.6\% \\
LB     &-           &101.3\%        &102.7\%  & 102.2\% \\

\bottomrule
\end{tabular}
\end{table}

\subsection{Complexity Comparison}
To evaluate the computational complexity of our scheme, \tabref{tab:timecomparison} compares its encoding time with the default HM encoder (i.e., JCTVC-M0257 for AI and JCTVC-M0036 for LB).
The results show that
our scheme and the other three competitors have brought only a small increase in encoding time.
In addition, compared with the other two SSIM-based schemes (i.e., Gao\cite{Gao2016} and Zhou\cite{Zhou2019}), the time increase of our scheme is slightly smaller.

\section{Conclusion}
In this study, the SSIM-based OBA and SSIM-based RDO are unified in our SOSR scheme.
Compared with the conventional SOMR scheme that is based on SSIM-based OBA and MSE-based RDO,
SOSR solves two problems.
First,
SSIM-based RDO is achieved based on the R-$\dssim$ cost with a $\lssim$-related Lagrangian multiplier, where the $\lssim$ is determined by the SSIM-based OBA.
In this way,
both OBA and RDO are optimized based on the unified R-$\dssim$ model without increasing the computation load.
Secondly,
by proposing to use $\lssim$ to control the RDO process,
there exists a causal relationship between $\lssim$ and the R-$\dssim$ generated by the encoding.
Accordingly, the joint relationship of R-$\dssim$-$\lssim$ can be exploited to calculate the R-$\dssim$ model parameters accurately.
Experimental results have validated that
by solving the two problems of SOMR, the proposed SOSR has achieved significant improvement in the R-$\dssim$ performance.
Moreover, compared with other state-of-the-art studies, SOSR also has an outstanding R-$\dssim$ performance in various configurations.

\bibliography{refs}
\bibliographystyle{IEEEtran}

\end{document}